\documentclass{llncs}
\usepackage{amsmath,amssymb,calc,ifthen}
\usepackage{float}
\usepackage[table,usenames,dvipsnames]{xcolor} 
\usepackage{tikz}
\usepackage{hyperref}
\usepackage{url}
\hypersetup{
colorlinks,
citecolor=blue,
filecolor=blue,
linkcolor=blue,
urlcolor=blue
}
\usepackage{makecell}
\usetikzlibrary{plotmarks,shapes}
\usepackage{amsmath,graphicx}
\usepackage{epstopdf}
\usepackage{subcaption}
\usepackage{graphicx}
\usepackage{color}

\usepackage{listings}
\usepackage{pdfpages}
\usepackage{enumitem} 

\usepackage[linesnumbered,noend]{algorithm2e}

\SetCommentSty{mycommfont}

\usepackage{filecontents}

\begin{document}

\definecolor{blue3}{HTML}{86B7FC} 
\definecolor{blue1}{HTML}{B5F1FF} 
\definecolor{blue2}{HTML}{E0F9FF} 

\title{TADPOLE Challenge: Accurate Alzheimer's disease prediction through crowdsourced forecasting of future data}
\titlerunning{TADPOLE Challenge: Accurate Alzheimer's disease prediction through crowdsourced forecasting of future data}

%

%
%

\author{R\u{a}zvan V. Marinescu\inst{1,2} \and Neil P. Oxtoby\inst{2} \and Alexandra L. Young\inst{2} \and Esther E. Bron\inst{3} \and Arthur W. Toga\inst{4} \and Michael W. Weiner\inst{5} \and Frederik Barkhof\inst{3,6} \and Nick C. Fox\inst{7} \and Polina Golland\inst{1} \and Stefan Klein\inst{3} \and Daniel C. Alexander\inst{2}} 



\institute{
Computer Science and Artificial Intelligence Laboratory, MIT, USA
\and
Centre for Medical Image Computing, University College London, UK
\and
Biomedical Imaging Group Rotterdam, Erasmus MC, Netherlands
\and
Laboratory of NeuroImaging, University of Southern California, USA
\and
Center for Imaging of Neurodegenerative Diseases, UCSF, USA
\and
Department of Radiology and Nuclear Medicine, VU Medical Centre, Netherlands\\
\and
Dementia Research Centre, UCL Institute of Neurology, UK
\email{tadpole@cs.ucl.ac.uk}
}

\maketitle              

\newcommand{\expFld}{.}

\begin{abstract}
The Alzheimer's Disease Prediction Of Longitudinal Evolution (TADPOLE) Challenge compares the performance of algorithms at predicting the future evolution of individuals at risk of Alzheimer's disease. TADPOLE Challenge participants train their models and algorithms on historical data from the Alzheimer's Disease Neuroimaging Initiative (ADNI) study. Participants are then required to make forecasts of three key outcomes for ADNI-3 rollover participants: clinical diagnosis, Alzheimer's Disease Assessment Scale Cognitive Subdomain (ADAS-Cog 13), and total volume of the ventricles -- which are then compared with future measurements. Strong points of the challenge are that the test data did not exist at the time of forecasting (it was acquired afterwards), and that it focuses on the challenging problem of cohort selection for clinical trials by identifying fast progressors. The submission phase of TADPOLE was open until 15 November 2017; since then data has been acquired until April 2019 from 219 subjects with 223 clinical visits and 150 Magnetic Resonance Imaging (MRI) scans, which was used for the evaluation of the participants' predictions. Thirty-three teams participated with a total of 92 submissions. No single submission was best at predicting all three outcomes. For diagnosis prediction, the best forecast (team \emph{Frog}), which was based on gradient boosting, obtained a multiclass area under the receiver-operating curve (MAUC) of 0.931, while for ventricle prediction the best forecast (team \emph{EMC1}), which was based on disease progression modelling and spline regression, obtained mean absolute error of 0.41\% of total intracranial volume (ICV). For ADAS-Cog 13, no forecast was considerably better than the benchmark mixed effects model (\emph{BenchmarkME}), provided to participants before the submission deadline. Further analysis can help understand which input features and algorithms are most suitable for Alzheimer's disease prediction and for aiding patient stratification in clinical trials. The submission system remains open via the website: \url{https://tadpole.grand-challenge.org/}

\keywords{Alzheimer's Disease, Future prediction, Community Challenge}
\end{abstract}

\section{Introduction}
\label{intro}

Accurate prediction of the onset of Alzheimer's disease (AD) and its longitudinal progression is important for care planning and for patient selection in clinical trials. Early detection will be critical in the successful administration of disease modifying treatments during presymptomatic phases of the disease prior to widespread brain damage, i.e. when pathological amyloid and tau accumulate \cite{mehta2017trials}. Moreover, accurate prediction of the evolution of subjects at risk of Alzheimer's disease will help to select homogeneous patient groups for clinical trials, thus reducing variability in outcome measures that can obscure positive effects on subgroups of patients who were at the right stage to benefit.

\begin{figure*}
 \centering
 \includegraphics[width=0.85\textwidth]{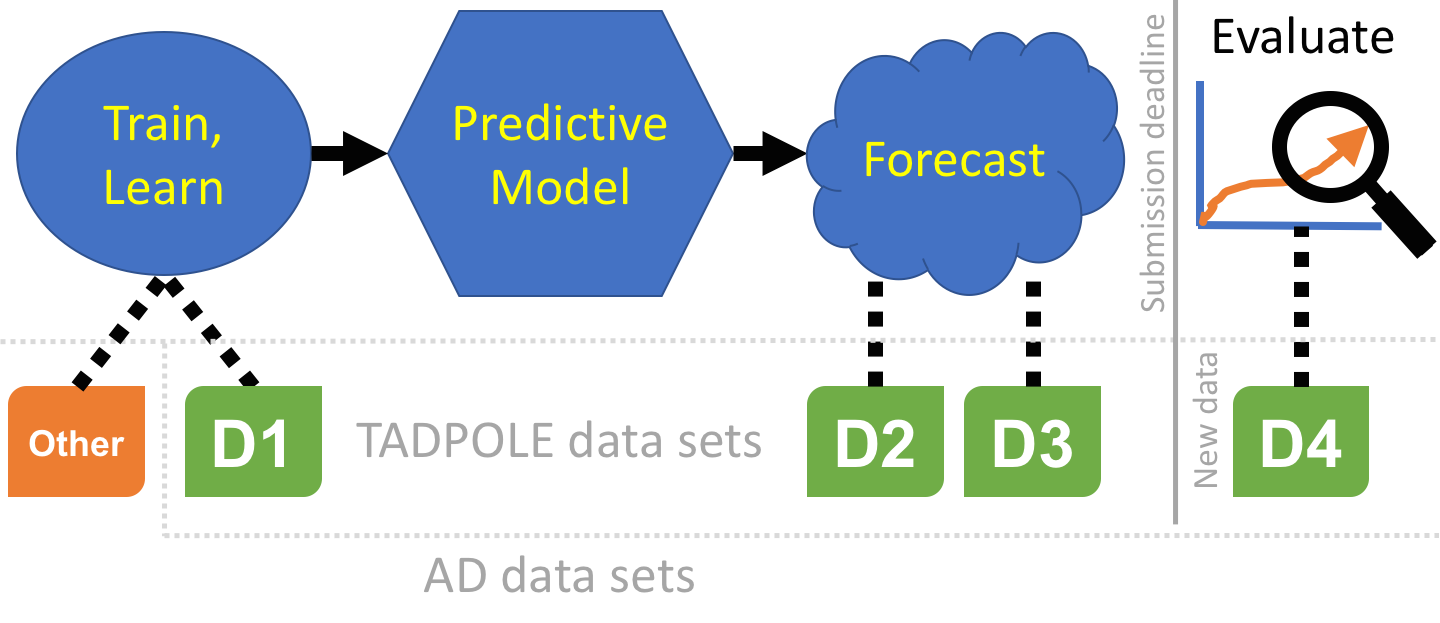}
 \caption{TADPOLE Challenge design. Participants are required to train a predictive model on a training dataset (D1 and/or others) and make forecasts for different datasets (D2, D3) by the submission deadline. Evaluation will be performed on a test dataset (D4) that is acquired after the submission deadline.}
 \label{fig:design}
 \end{figure*}

Several approaches for predicting AD-related target variables (e.g. clinical diagnosis, cognitive/imaging biomarkers) have been proposed which leverage multimodal biomarker data available in AD. Traditional longitudinal approaches based on statistical regression model the relationship of the target variables with other known variables, such as clinical diagnosis \cite{scahill2002mapping}, cognitive test scores \cite{yang2011quantifying}, or time to conversion between diagnoses \cite{guerrero2016instantiated}. Another approach involves supervised machine learning techniques such as support vector machines, random forests, and artificial neural networks, which use pattern recognition to learn the relationship between the values of a set of predictors (biomarkers) and their labels (diagnoses). These approaches have been used to discriminate AD patients from cognitively normal individuals \cite{kloppel2008automatic}, and for discriminating at-risk individuals who convert to AD in a certain time frame from those who do not \cite{young2013accurate}. The emerging approach of disease progression modelling \cite{young2014data,lorenzi2017probabilistic} aims to reconstruct biomarker trajectories or other disease signatures across the disease progression timeline, without relying on clinical diagnoses or estimates of time to symptom onset. Such models show promise for predicting AD biomarker progression at group and individual levels. However, previous evaluations within individual publications are not systematic and reliable because: (1) they use different data sets or subsets of the same dataset, different processing pipelines and different evaluation metrics and (2) over-training can occur due to heavy use of popular training datasets. Currently we lack a comprehensive comparison of the capabilities of these methods on standardised tasks relevant to real-world applications.

Community challenges have consistently proven effective in moving forward the state-of-the-art in technology to address specific data-analysis problems by providing platforms for unbiased comparative evaluation and incentives to maximise performance on key tasks. For Alzheimer's disease prediction in particular, previous challenges include the CADDementia challenge \cite{bron2015standardized} which aimed to identify clinical diagnosis from MRI scans. A similar challenge, the  ``International challenge for automated prediction of MCI from MRI data`` \cite{sarica2018editorial} asked participants to predict diagnosis and conversion status from extracted MRI features of subjects from the ADNI study \cite{weiner2017recent}. Yet another challenge, The Alzheimer's Disease Big Data DREAM Challenge \cite{allen2016crowdsourced}, asked participants to predict cognitive decline from genetic and MRI data.  However, most of these challenges have not evaluated the ability of algorithms to predict clinical diagnosis and other biomarkers at future timepoints and largely used training data from a limited set of modalities. The one challenge that asked participants to estimate a biomarker at future timepoints (cognitive decline in one of the DREAM sub-challenges) used only genetic and cognitive data for training, and aimed to find genetic loci that could predict cognitive decline. Therefore, standardised evaluation of algorithms needs to be done on biomarker prediction at future timepoints, with the aim of improving clinical trials through enhanced patient stratification.

The Alzheimer's Disease Prediction Of Longitudinal Evolution (TADPOLE) Challenge aims to identify the data, features and approaches that are most predictive of future progression of subjects at risk of AD. The challenge focuses on forecasting the evolution of three key AD-related domains: clinical diagnosis, cognitive decline, and neurodegeneration (brain atrophy). In contrast to previous challenges, our challenge is designed to inform clinical trials through identification of patients most likely to benefit from an effective treatment, i.e., those at early stages of disease who are likely to progress over the short-to-medium term (defined as 1-5 years). Since the test data did not exist at the time of forecast submissions, the challenge provides a performance comparison substantially less susceptible to many forms of potential bias than previous studies and challenges. The design choices were published \cite{marinescu2018tadpole} before the test set was acquired and analysed. TADPOLE also goes beyond previous challenges by drawing on a vast set of multimodal measurements from ADNI which support prediction of AD progression. 

This article presents the design of the TADPOLE Challenge and outlines preliminary results.

\section{Competition Design}
\label{design}

The aim of TADPOLE is to predict future outcome measurements of subjects at-risk of AD, enrolled in the ADNI study. A history of informative measurements from ADNI (imaging, psychology, demographics, genetics, etc.) from each individual is available to inform forecasts. TADPOLE participants were required to predict future measurements from these individuals and submit their predictions before a given submission deadline. Evaluation of these forecasts occurred post-deadline, after the measurements had been acquired. A diagram of the TADPOLE flow is shown in Fig \ref{fig:design}.


TADPOLE challenge participants were required to make month-by-month forecasts of three key biomarkers: (1) clinical diagnosis which is either cognitively normal (CN), mild cognitive impairment (MCI) or probable Alzheimer's disease (AD); (2) Alzheimer's Disease Assessment Scale Cognitive Subdomain (ADAS-Cog 13) score; and (3) ventricle volume (divided by intra-cranial volume). TADPOLE forecasts are required to be probabilistic and some evaluation metrics will account for forecast probabilities provided by participants. 

\section{ADNI data aggregation and processing}

TADPOLE Challenge organisers provided participants with a standard ADNI-derived dataset (available via the Laboratory Of NeuroImaging data archive at \url{adni.loni.usc.edu}) to train algorithms, removing the need for participants to pre-process the ADNI data or merge different spreadsheets. Software code used to generate the standard datasets is openly available on Github\footnote{https://github.com/noxtoby/TADPOLE}. The challenge data includes: (1) CSF markers of amyloid-beta and tau deposition; (2) various imaging modalities such as magnetic resonance imaging (MRI), positron emission tomography (PET) using several tracers: FDG (hypometabolism), AV45 (amyloid), AV1451 (tau) as well as diffusion tensor imaging (DTI); (3) cognitive assessments such as ADAS-Cog 13 acquired in the presence of a clinical expert; (4) genetic information such as alipoprotein E4 (APOE4) status extracted from DNA samples; and (5) general demographic information such as age and gender. Extracted features from this data were merged into a final spreadsheet and made available online.

The imaging data was pre-processed with standard ADNI pipelines.  For MRI scans, this included correction for gradient non-linearity, B1 non-uniformity correction and peak sharpening\footnote{see \url{http://adni.loni.usc.edu/methods/mri-analysis/
mri-pre-processing}}. Meaningful regional features such as volume and cortical thickness were extracted using Freesurfer. Each PET image (FDG, AV45, AV1451), which consists of a series of dynamic frames, had its frames co-registered, averaged across the dynamic range, standardised with respect to the orientation and voxel size, and smoothed to produce a uniform resolution of 8mm full-width/half-max (FWHM)\footnote{see \url{http://adni.loni.usc.edu/methods/pet-analysis/pre-processing}}. Standardised uptake value ratio (SUVR) measures for relevant regions-of-interest were extracted after registering the PET images to corresponding MR images using SPM5. DTI scans were corrected for head motion and eddy-current distortion, skull-stripped, EPI-corrected, and finally aligned to the T1 scans. Diffusion tensor summary measures were estimated based on the Eve white-matter atlas. 

\subsection{TADPOLE Datasets}
\label{datasets}

In order to evaluate the effect of different methodological choices, we prepared four ``standard`` data sets: the \underline{\smash{D1 standard training set}} contains longitudinal data from the entire ADNI history; the  \underline{\smash{D2 longitudinal prediction set}} contains all available data from the ADNI rollover individuals, for whom challenge participants are asked to provide forecasts; the \underline{\smash{D3 cross-sectional prediction set}} contains a single (most recent) time point and a limited set of variables from each rollover individual -- this represents the information typically available in a clinical trial; the              
\underline{\smash{D4 test set}} contains visits from ADNI rollover subjects after 1 Jan 2018, which contain at least one of the following: diagnostic status, ADAS score, or ventricle volume from MRI -- this dataset did not exist at the time of submitting forecasts. Full demographics for D1--D4 are given in Table \ref{tab:biomk_data_available}.

\begin{table}
\centering
\begin{tabular}{c|cccc}

\textbf{Measure} &            \textbf{D1} &            \textbf{D2} &           \textbf{D3} &           \textbf{D4} \\
Subjects &          1667 &           896 &          896 &          219 \\
\hline
& \multicolumn{4}{c}{\textbf{Cognitively Normal} }\\
Number (\% total) &   508 (30\%) &   369 (41\%) &  299 (33\%) &   94 (42\%) \\
Visits per subject &     8.3 $\pm$ 4.5 &     8.5 $\pm$ 4.9 &    1.0  $\pm$ 0.0 &    1.0  $\pm$ 0.2 \\
Age &    74.3  $\pm$ 5.8 &    73.6 $\pm$ 5.7 &   72.3 $\pm$ 6.2 &   78.4 $\pm$ 7.0 \\
Gender (\% male) &         48\% &         47\% &        43\% &        47\% \\
MMSE &    29.1  $\pm$ 1.1 &    29.0  $\pm$ 1.2 &   28.9  $\pm$ 1.4 &   29.1 $\pm$ 1.1 \\
Converters (\% total CN) &     18 (3.5\%) &      9 (2.4\%) &  - & - \\
\hline
& \multicolumn{4}{c}{\textbf{Mild Cognitive Impairment} }\\
Number (\% total) &   841 (50.4\%) &   458 (51.1\%) &  269 (30.0\%) &   90 (41.1\%) \\
Visits per subject &     8.2 $\pm$ 3.7 &     9.1 $\pm$ 3.6 &    1.0 $\pm$ 0.0 &    1.1 $\pm$ 0.3 \\
Age &    73.0 $\pm$ 7.5 &    71.6 $\pm$ 7.2 &   71.9 $\pm$ 7.1 &   79.4 $\pm$ 7.0 \\
Gender (\% male) &         59.3\% &         56.3\% &        58.0\% &        64.4\% \\
MMSE &    27.6 $\pm$ 1.8 &    28.0 $\pm$ 1.7 &   27.6 $\pm$ 2.2 &   28.1 $\pm$ 2.1 \\
Converters (\% total MCI) &   117 (13.9\%) &     37 (8.1\%) &      -        &    9 (10.0\%) \\
\hline
& \multicolumn{4}{c}{\textbf{Alzheimer's Disease}} \\
Number (\% total) &   318 (19.1\%) &     69 (7.7\%) &  136 (15.2\%) &   29 (13.2\%) \\
Visits per subject &     4.9 $\pm$ 1.6 &     5.2 $\pm$ 2.6 &    1.0 $\pm$ 0.0 &    1.1 $\pm$ 0.3 \\
Age &    74.8 $\pm$ 7.7 &    75.1 $\pm$ 8.4 &   72.8 $\pm$ 7.1 &   82.2 $\pm$ 7.6 \\
Gender (\% male) &         55.3\% &         68.1\% &        55.9\% &        51.7\% \\
MMSE &    23.3 $\pm$ 2.0 &    23.1 $\pm$ 2.0 &   20.5 $\pm$ 5.9 &   19.4 $\pm$ 7.2 \\
Converters (\% total AD) &       -        &      -         &        -      &    9 (31.0\%) \\
                
 \hline
\end{tabular}

 \caption{Summary of TADPOLE datasets D1--D4. Each subject has been allocated to either Cognitively Normal, MCI or AD group based on diagnosis at the first available visit within each dataset.}
 \label{tab:biomk_data_available}
\end{table}

\section{Submissions and evaluation}
\label{submissions}

The challenge had a total of 33 participating teams, who submitted a total of 58 forecasts from D2, 34 forecasts from D3, and 6 forecasts from custom prediction sets. Table \ref{tab:submissions_desc} summarises the top-3 winner methods in terms of input features used, handling of missing data and predictive models: \emph{Frog} used a gradient boosting method, which combined many weak predictors to build a strong predictor; \emph{EMC1} derived a ``disease state`` variable aggregating multiple features together and then used an SVM and 2D splines for prediction, while VikingAI used a latent-time parametric model with subject- and feature-specific parameters -- see \cite{tadpole2019results} for full method details. We also describe three benchmark models which were provided to participants at the start of the challenge: (i) \emph{BenchmarkLastVisit} uses the measurement at the last available visit, (ii) \emph{BenchmarkME-APOE} uses a mixed effects model with APOE status as covariate and (iii) BenchmarkSVM uses an out-of-the-box support vector machine (SVM) and regressor for forecast. 

\begin{table}
 \centering
 \fontsize{8}{11}\selectfont
 \begin{tabular}{c | >{\centering\arraybackslash}p{1.3cm} >{\centering\arraybackslash}p{1.2cm} >{\centering\arraybackslash}p{2cm} >{\centering\arraybackslash}p{2cm} >{\centering\arraybackslash}p{2cm}}
\textbf{Submission} & \textbf{Extra$^{\dagger}$ Features}  & \textbf{Nr. of features} & \textbf{Missing data imputation} & \textbf{Diagnosis prediction} & \textbf{ADAS/Vent. prediction}\\
\Xhline{2.5\arrayrulewidth}
Frog & most features & 70+420* & none & gradient boosting & gradient boosting \\ 
\hline
EMC1-Std & MRI, ASL, cognitive & 250 & nearest neighbour & DPM SVM 2D-spline  & DPM 2D-spline\\
\hline
VikingAI-Sigmoid & MRI, cognitive, tau & 10 & none & DPM + ordered logit & DPM \\ 
\hline
BenchmarkLastVisit & - & 3 & none & constant model & constant model \\
\hline
BenchmarkME-APOE & APOE & 4 & none & Gaussian model & linear mixed effects model \\ 
\hline
BenchmarkSVM & age, APOE & 6 & mean of previous values & SVM & support vector regressor\\
 \end{tabular}
 \caption{Summary of benchmarks and top-3 methods used in the TADPOLE submissions. DPM -- disease progression model. ($^{\dagger}$) Aside from the three target biomarkers (*) Augmented features: e.g. min/max, trends, moments.}
 \label{tab:submissions_desc}
\end{table}

For evaluation of clinical status predictions, we used similar metrics to those that proved effective in the CADDementia challenge \cite{bron2015standardized}: (i) the multiclass area under the receiver operating curve (MAUC); and (ii) the overall balanced classification accuracy (BCA). For ADAS and ventricle volume, we used three metrics: (i) mean absolute error (MAE), (ii) weighted error score (WES) and (iii) coverage probability accuracy (CPA). BCA and MAE focus purely on prediction accuracy ignoring confidence, MAUC and WES include confidence, while CPA provides an assessment of the confidence interval only. Complete formulations for these can be found in Table \ref{tab:perfMetrics}, with detailed explanations in the TADPOLE design paper \cite{marinescu2018tadpole}. To compute an overall rank, we first calculated the sum of ranks from MAUC, ADAS MAE and Ventricle MAE for each submission, and the overall ranking was derived from these sums of ranks.

\begin{table}
 \centering
 \fontsize{8}{12}\selectfont
 \begin{tabular}{>{\centering\arraybackslash}m{4.1cm} | >{\centering\arraybackslash}m{7.8cm}}
\textbf{Formula} & \textbf{Definitions}\\
\Xhline{2.5\arrayrulewidth}
$mAUC =\frac{2}{L(L-1)}\sum_{i=2}^L\sum_{j=1}^{i}\hat{A}(c_i,c_j)$ & $n_i$, $n_j$ -- number of points from class $i$ and $j$. $S_{ij}$ -- the sum of the ranks of the class $i$ test points, after ranking all the class $i$ and $j$ data points in increasing likelihood of belonging to class $i$, $L$ -- number of data points \\
\hline
$BCA = \frac{1}{2L}\sum_{i=1}^L \left[\frac{TP}{TP+FN}+\frac{TN}{TN+FP}\right]$ & $TP_i$, $FP_i$, $TN_i$, $FN_i$ – the number of true positives, false positives, true negatives and false negatives for class $i$
$L$ – number of data points \\
\hline
$MAE = \frac{1}{N}\sum_{i=1}^{N}\left|{\tilde{M}_i-M_i}\right|$ & $M_i$ is the actual value in individual $i$ in future data. $\tilde{M}_i$ is the participant's best guess at $M_i$ and $N$ is the number of data points \\
\hline
$WES =\frac{\sum_{i=1}^{N}\tilde{C}_i\left|\tilde{M}_i-M_i\right|}{\sum_{i=1}^{N}\tilde{C}_i}$ & $M_i$, $\tilde{M}_i$ and $N$ defined as above. 
$\tilde{C}_i = (C_{+} - C_{-})^{-1}$, where $\left[C_{-}, C_{+}\right]$ is the 50\% confidence interval\\
\hline
$CPA = |ACP - 0.5|$ & actual coverage probability (ACP) - the proportion of measurements that fall within the 50\% confidence interval.\\
 \end{tabular}
 \caption{TADPOLE performance metric formulas and definitions for the terms.}
 \label{tab:perfMetrics}
\end{table}


\section{Results}

While full results can be found on the TADPOLE website \cite{tadpole2019results}, here we only include the top-3 winners. Table \ref{tab:resTable} compiles all metrics for top-3 TADPOLE forecasts from the D2 prediction set. The best overall performance was obtained by team \emph{Frog}, with a clinical diagnosis MAUC of 0.931, ADAS MAE of 4.85 and Ventricle MAE of 0.45. Among the benchmark methods, \emph{BenchmarkME-APOE} had the best overall rank of 18, obtaining an MAUC of 0.82, ADAS MAE of 4.75 and Ventricle MAE of 0.57. In terms of diagnosis predictions, \emph{Frog} had an overall MAUC score of 0.931. For ADAS prediction, \emph{BenchmarkME-APOE} had the best MAE of 4.75. For Ventricle prediction, \emph{EMC1-Std} had the best MAE of 0.41 and WES of 0.29. In terms of the most accurate confidence interval estimates, \emph{VikingAI} achieved the best CPA scores of 0.02 for ADAS and 0.2 for Ventricles. 

\begin{table}
\centering
\fontsize{8}{12}\selectfont
\begin{tabular}{cc|cc|ccc|ccc}
& \textbf{Overall} & \multicolumn{2}{c|}{\textbf{Diagnosis}} & \multicolumn{3}{c|}{\textbf{ADAS}}  & \multicolumn{3}{c}{\textbf{Ventricles (\% ICV)}}\\
\textbf{Submission} & \textbf{Rank} & \textbf{MAUC} & \textbf{BCA} & \textbf{MAE} & \textbf{WES} & \textbf{CPA} & \textbf{MAE} & \textbf{WES} & \textbf{CPA}\\
\hline
Frog & 1 & 0.931 & 0.849 & 4.85 & 4.74 & 0.44 & 0.45 & 0.33 & 0.47\\ 
EMC1-Std & 2 & 0.898 & 0.811 & 6.05 & 5.40 & 0.45 &  0.41 & 0.29 & 0.43\\ 
VikingAI-Sigmoid & 3 & 0.875 & 0.760 & 5.20 & 5.11 & 0.02 & 0.45 & 0.35 & 0.20\\ 
BenchmarkME-APOE & 18 & 0.822 & 0.749 & 4.75 & 4.75 & 0.36 & 0.57 & 0.57 & 0.40\\ 
BenchmarkSVM & 34 & 0.836 & 0.764 & 6.82 & 6.82 & 0.42 & 0.86 & 0.84 & 0.50\\
BenchmarkLastVisit & 40 & 0.774 & 0.792 & 7.05 & 7.05 & 0.45 & 0.63 & 0.61 & 0.47\\
\end{tabular}
\vspace{0.5em}
\caption{Ranked forecasting scores for benchmark models and top-3 TADPOLE submissions. }
\label{tab:resTable}
\end{table}

\section{Discussion}


In the current work we have outlined the design and key results of TADPOLE Challenge, which aims to identify algorithms and features that can best predict the evolution of Alzheimer's disease. Despite the small number of converters in the training set, the methods were able to accurately forecast the clinical diagnosis and ventricle volume, although they found it harder to forecast cognitive test scores. Compared to the benchmark models, the best submissions had considerably smaller errors that represented only a small fraction of the errors obtained by benchmark models (0.42 for clinical diagnosis MAUC and 0.71 for ventricle volume MAE). For clinical diagnosis, this suggests that more than half of the subjects originally misdiagnosed by the best benchmark model (\emph{BenchmarkSVM}) are now correctly diagnosed with the new methods. Moreover, the results suggest that we do not have a clear winner on all categories. While team Frog had the best overall submission with the lowest sum of ranks, for each performance metric individually we had different winners.

Additional work currently in progress \cite{tadpole2019results} suggests that consensus methods based on averaging predictions from all participants perform better than any single individual method. This demonstrates the power of TADPOLE in achieving state-of-the-art prediction accuracy through crowd-sourcing prediction models.

The TADPOLE Challenge and its preliminary results presented here are of importance for the design of future clinical trials and more generally may be applicable to a clinical setting. The best algorithms identified here could be used for subject selection or stratification in clinical trials, e.g. by enriching trial inclusion with fast progressors to increase the statistical power to detect treatment changes. Alternatively, a stratification could be implemented based on predicted ``fast progressors``  and ``slow progressors`` to reduce imbalances between arms. In order to make these models applicable to clinical settings, application in a clinical sample should be tested outside ADNI and further validation in a subject population with post-mortem confirmation would be desirable, as clinical diagnosis of probable AD only has moderate agreement with gold-standard neuropathological post-mortem diagnosis (70.9\% -- 87.3\% sensitivity and 44.3\% -- 70.8\% specificity, according to \cite{beach2012accuracy}). We hope such a validation will be possible in the future, with the advent of neuropathological confirmation in large, longitudinal, multimodal datasets such as ADNI.

In future work, we plan to analyse which features and methods were most useful for predicting AD progression, and assess if the results are sufficient to improve stratification for AD clinical trials. We also plan to evaluate the impact and interest of the first phase of TADPOLE within the community, to guide decisions on whether to organise further submission and evaluation phases.


\section{Acknowledgements}

TADPOLE Challenge has been organised by the European Progression Of Neurological Disease (EuroPOND) consortium, in collaboration with the ADNI. We thank all the participants and advisors, in particular Clifford R. Jack Jr. from Mayo Clinic, Rochester, United States and Bruno M. Jedynak from Portland State University, Portland, United States for useful input and feedback.

The organisers are extremely grateful to The Alzheimer's Association, The Alzheimer's Society and Alzheimer's Research UK for sponsoring the challenge by providing the prize fund and providing invaluable advice into its construction and organisation. Similarly, we thank the ADNI leadership and members of our advisory board and other members of the EuroPOND consortium for their valuable advice and support.

RVM was supported by the EPSRC Centre For Doctoral Training in Medical Imaging with grant EP/L016478/1 and by the Neuroimaging Analysis Center with grant NIH NIBIB NAC P41EB015902. NPO, FB, SK, and DCA are supported by EuroPOND, which is an EU Horizon 2020 project. ALY was supported by an EPSRC Doctoral Prize fellowship and by EPSRC grant EP/J020990/01. PG was supported by NIH grant NIBIB NAC P41EB015902 and by grant NINDS R01NS086905. DCA was supported by EPSRC grants J020990, M006093 and M020533. The UCL-affiliated researchers received support from the NIHR UCLH Biomedical Research Centre. Data collection and sharing for this project was funded by the Alzheimer's Disease Neuroimaging Initiative (ADNI) (National Institutes of Health Grant U01 AG024904) and DOD ADNI (Department of Defense award number W81XWH-12-2-0012). FB was supported by the NIHR UCLH Biomedical Research Centre and the AMYPAD project, which has received support from the EU-EFPIA Innovative Medicines Initiatives 2 Joint Undertaking (AMYPAD project, grant 115952). This project has received funding from the European Union Horizon 2020 research and innovation programme under grant agreement No 666992.

\bibliographystyle{splncs}
\bibliography{bibliography}

\end{document}